# A novel hybrid protocol for semiquantum key distribution and semiquantum secret sharing

Tian-Yu Ye*, Xiao Tan

College of Information & Electronic Engineering, Zhejiang Gongshang University, Hangzhou 310018, P.R.China
E-mail：yetianyu@mail.zjgsu.edu.cn

**Abstract:** In this paper, a novel hybrid protocol for semiquantum key distribution (SQKD) and semiquantum secret sharing (SQSS) was constructed by using GHZ-like states. This protocol is capable of establishing two different private keys between one quantum party and two semiquantum parties respectively, and making two semiquantum parties share another private key of the quantum party in the meanwhile. The usages of delay lines, Pauli operations, Hadamard gates and quantum entanglement swapping are not required. Moreover, the semiquantum parties are not necessary to be equipped with any quantum memory. We validate in detail that this protocol resists various attacks from Eve, including the Trojan horse attacks, the entangle-measure attack, the double controlled-not (CNOT) attacks, the measure-resend attack and the intercept-resend attack. To our best knowledge, this protocol is the only protocol which possesses the functions of both SQKD and SQSS simultaneously until now.

## 1 Introduction

It is popularly known that quantum cryptography was derived from BB84 quantum key distribution (QKD) protocol of Bennett and Brassard. Because quantum cryptography can attain the unconditional security theoretically through the laws of quantum mechanics, many enthusiasms have been put onto studying it since its birth. Consequently, numerous interesting and important branches of quantum cryptography have been created, such as QKD [1-4], quantum secret sharing (QSS) [5-11] and so on. Naturally, different branches have different functions. The function of QKD is to establish a private key between two remote lawful parties; and the function of QSS is to make two lawful parties share the private key of the third party on the condition that only when they cooperate can they recover it.

In order to cut down the burden of quantum state preparation and quantum measurement for a partial parties, Boyer *et al.* [12,13] put forward a brand-new kind of quantum cryptography named as semiquantum cryptography in 2007. In a semiquantum cryptography scheme, the semiquantum party is restricted within limited quantum operations, such as producing particles in the classical basis $\{|0\rangle,|1\rangle\}$, measuring particles in the classical basis $\{|0\rangle,|1\rangle\}$, transmitting particles and reordering particles. Subsequently, many attentions have been absorbed into studying semiquantum cryptography. Consequently, many interesting and important branches of semiquantum cryptography have been established, such as semiquantum key distribution (SQKD) [12-26], semiquantum secret sharing (SQSS) [27-35] and so on. Apparently, each of the protocols of Refs.[12-26] only has the function of SQKD, while each of the protocols of Refs.[27-35] only has the function of SQSS. Unfortunately, at present, there is no protocol which simultaneously possesses the functions of both SQKD and SQSS.

Refs.[22,24] use GHZ-like states to design SQKD protocols, while Refs.[27,31] use GHZ-like states to design SQSS protocols. Ref.[30] put forward a new shadow key construction strategy to enhance the qubit efficiency for SQSS protocols. Inspired by Refs.[22,24,27,30,31], in this paper, we are devoted to utilizing GHZ-like states to design a novel hybrid protocol for SQKD and SQSS. Concretely speaking, the proposed protocol is capable of establishing two different private keys between one quantum party and two semiquantum parties respectively, and making two semiquantum parties share another private key of the quantum party in the meanwhile. To our best knowledge, the proposed protocol is the only protocol which possesses the functions of both SQKD and SQSS simultaneously until now.

## 2 The proposed hybrid protocol for SQKD and SQSS

Eight GHZ-like states compose an orthonormal basis (i.e., the GHZ-like basis) for the space of a tripartite quantum system, which can be represented as [36]

$$|G_0\rangle = \frac{1}{2}(|000\rangle + |011\rangle + |101\rangle + |110\rangle), |G_1\rangle = \frac{1}{2}(|001\rangle + |010\rangle + |100\rangle + |111\rangle),$$

$$|G_2\rangle = \frac{1}{2}(|000\rangle - |011\rangle - |101\rangle + |110\rangle), |G_3\rangle = \frac{1}{2}(|001\rangle - |010\rangle - |100\rangle + |111\rangle),$$

$$|G_4\rangle = \frac{1}{2}(|000\rangle - |011\rangle + |101\rangle - |110\rangle), |G_5\rangle = \frac{1}{2}(|001\rangle - |010\rangle + |100\rangle - |111\rangle),$$

$$|G_6\rangle = \frac{1}{2}(|000\rangle + |011\rangle - |101\rangle - |110\rangle), |G_7\rangle = \frac{1}{2}(|001\rangle + |010\rangle - |100\rangle - |111\rangle). \quad (1)$$

$|G_1\rangle$ can further be denoted as

$$|G_1\rangle_{abc} = \frac{1}{\sqrt{2}}\left(|0\rangle_a |\psi^+\rangle_{bc} + |1\rangle_a |\phi^+\rangle_{bc}\right)$$

$$= \frac{1}{\sqrt{2}}\left(|\psi^+\rangle_{ab} |0\rangle_c + |\phi^+\rangle_{ab} |1\rangle_c\right)$$

$$= \frac{1}{\sqrt{2}}\left(|\psi^+\rangle_{ac} |0\rangle_b + |\phi^+\rangle_{ac} |1\rangle_b\right). \quad (2)$$

where $|\psi^+\rangle = \frac{1}{\sqrt{2}}(|01\rangle + |10\rangle)$ and $|\phi^+\rangle = \frac{1}{\sqrt{2}}(|00\rangle + |11\rangle)$.

There are three communicants, Alice, Bob and Charlie. Here, Alice possesses unlimited quantum capabilities, while Bob and Charlie only have limited quantum capabilities. In the following, we construct a hybrid protocol of SQKD and SQSS. In this hybrid protocol, Alice can distribute two different private keys to Bob and Charlie simultaneously, and in the meanwhile, Bob and Charlie can recover another private key of Alice only when they cooperate together. Here, $|0\rangle$ is coded into the classical bit 0, while $|1\rangle$ is coded into the classical bit 1.

**Step 1:** Alice prepares $8(n+\tau)$ $|G_1\rangle_{abc}$, where $n$ is the length of private key and $\tau$ is a fixed parameter greater than 0. Afterward, Alice classifies these quantum states into three sequences: $S_l = \{S_l^1, S_l^2, \cdots, S_l^{8(n+\tau)}\}$, where $S_l^t$ is the particle with subscript $l$ of the $t$ th $|G_1\rangle_{abc}$, $l = a,b,c$ and $t = 1,2,\cdots,8(n+\tau)$. Alice keeps $S_a$ stationary, and sends $S_b$ to Bob and $S_c$ to Charlie. Note that the particles of $S_b$ and $S_c$ are sent out in a one-by-one manner. In other words, except the first particle, the next one is sent out by Alice only after the previous one has been obtained by her.

**Step 2:** Bob places a wavelength filter and a photon number splitter in front of his devices to eliminate the negative influence of the Trojan horse attacks from an eavesdropper. Bob randomly performs the MEASURE-RESEND mode or the REFLECT mode for each received particle of $S_b$.

Charlie also places a wavelength filter and a photon number splitter in front of his devices to eliminate the negative influence of the Trojan horse attacks from an eavesdropper. Charlie also randomly performs the MEASURE-RESEND mode or the REFLECT mode for each received particle of $S_c$. Here, the REFLECT mode means that the receiver returns the received particle back to the sender without disturbance, while the MEASURE-RESEND mode means that the receiver uses the $Z$ basis (i.e., $\{|0\rangle,|1\rangle\}$) to measure the received particle, generate a new one based on the measurement result and sends it to the sender.

**Step 3:** Alice announces that she has received all particles from Bob and Charlie. Then, she asks Bob and Charlie to inform her the positions where they selected the MEASURE-RESEND mode.

**Step 4:** Alice performs her actions according to Bob and Charlie's actions, just as described in Table 1.

Case ①: both Bob and Charlie entered into the MEASURE-RESEND mode. Alice imposes the $Z$ basis measurements on the particle from Bob, the particle from Charlie and the corresponding particle in her own hand. Alice randomly picks out half of this kind of positions, and asks Bob and Charlie to inform her their corresponding $Z$ basis measurements. If Eve is not online, for each selected position, Alice's $Z$ basis measurement results on the particle from Bob, the particle from Charlie and the corresponding particle in her own hand should be correctly related to Bob and Charlie's $Z$ basis measurement results, according to the definition of $|G_1\rangle_{abc}$ described in Eq.(1);

Case ②: Bob entered into the MEASURE-RESEND mode and Charlie entered into the REFLECT mode. Alice imposes the $Z$ basis measurement on the particle from Bob and the Bell basis measurement on the corresponding particle in her own hand and the particle from Charlie. Alice randomly picks out half of this kind of positions, and asks Bob to inform her his corresponding $Z$ basis measurements. If Eve is not online, for each selected position, Alice's $Z$ basis measurement result on the particle from Bob and the Bell basis measurement result on the corresponding particle in her own hand and the particle from Charlie should be correctly related to Bob's $Z$ basis measurement result, according to Eq.(2);

Case ③: Bob entered into the REFLECT mode and Charlie entered into the MEASURE-RESEND mode. Alice imposes the $Z$ basis measurement on the particle from Charlie and the Bell basis measurement on the corresponding particle in her own hand and the particle from Bob. Alice randomly picks out half of this kind of positions, and asks Charlie to inform her his corresponding $Z$ basis measurements. If Eve is not online, for each selected position, Alice's $Z$ basis measurement result on the particle from Charlie and the Bell basis measurement result on the corresponding particle in her own hand and the particle from Bob should be correctly related to Charlie's $Z$ basis measurement result, according to Eq.(2);

Case ④: both Bob and Charlie entered into the REFLECT mode. Alice imposes the GHZ-like basis measurement on the corresponding particle in Alice's hand, the particle from Bob and the particle from Charlie. If Eve is not online, for each position belonging to this Case, Alice's GHZ-like basis measurement result on the corresponding particle in Alice's hand, the particle from Bob and the particle from Charlie should be identical to $|G_1\rangle_{abc}$.

If either of the error rate of the above four Cases is unreasonably high, the communication will be halted and restarted from Step 1.

**Step 5:** After the above security check processes, there are approximately $n+\tau$ positions belonging to Case ② left which can be utilized to derive a private key between Alice and Bob.

Alice and Bob individually use their first $n$ remaining $Z$ basis measurement results belonging to Case ② to produce the private key $K_{AB}$ between them.

Table 1  The relations among Alice, Bob and Charlie's actions

| Case | Bob's action | Charlie's action | Alice's action |
|---|---|---|---|
| ① | MEASURE-RESEND | MEASURE-RESEND | Impose the $Z$ basis measurements on the particle from Bob, the particle from Charlie and the corresponding particle in Alice's hand |
| ② | MEASURE-RESEND | REFLECT | Impose the $Z$ basis measurement on the particle from Bob and the Bell basis measurement on the corresponding particle in Alice's hand and the particle from Charlie |
| ③ | REFLECT | MEASURE-RESEND | Impose the $Z$ basis measurement on the particle from Charlie and the Bell basis measurement on the corresponding particle in Alice's hand and the particle from Bob |
| ④ | REFLECT | REFLECT | Impose the GHZ-like basis measurement on the corresponding particle in Alice's hand, the particle from Bob and the particle from Charlie |

After the above security check processes, there are approximately $n+\tau$ positions belonging to Case ③ left which can be utilized to derive a private key between Alice and Charlie. Alice and Charlie individually use their first $n$ remaining $Z$ basis measurement results belonging to Case ③ to produce the private key $K_{AC}$ between them.

After the above security check processes, there are approximately $n+\tau$ positions belonging to Case ① left which can be utilized to derive a private key of Alice shared by Bob and Charlie together. Bob and Charlie individually use their first $n$ remaining $Z$ basis measurement results belonging to Case ① to produce the shadow keys $K_B$ and $K_C$. Alice uses her first $n$ remaining $Z$ basis measurement results on the particles in her hand belonging to Case ① to produce the private key $K_A$. According to Eq.(1), the relationship that $\overline{K}_A = K_B \oplus K_C$ stands. Here, $\oplus$ is the bitwise XOR operation, while $\overline{x}$ is the value derived from performing the bitwise NOT operation on $x$, where $x$ is a bit string. It is easy to understand that only when Bob and Charlie cooperate together can they recover $K_A$.

## 3  Security analysis

In this section, we analyze the security of this scheme against an external eavesdropper Eve.

(1) The Trojan horse attacks

In this protocol, not only Bob but also Charlie place a wavelength filter and a photon number splitter in front of his devices, hence the negative influence of the Trojan horse attacks from Eve can be erased, according to Refs.[37,38].

(2) The entangle-measure attack

Eve's entangle-measure attack can be depicted as Fig.1, where $U_E$ and $U_F$ are Eve's two unitaries, sharing a probe space with the initial state $|\chi\rangle_E$. Apparently, $U_E$ attacks the particles sent to Bob and Charlie, while $U_F$ attacks the particles sent out from Bob and Charlie. By virtue of the shared probe, Eve can attack the particles sent out from Bob and Charlie based on the knowledge acquired from $U_E$ [12,13].

**Theorem 1.** *Suppose that Eve's entangle-measure attack can be modelled with $U_E$ and $U_F$, sharing a probe space with the initial state $|\chi\rangle_E$. Here, $U_E$ attacks the particles sent to Bob and Charlie, while $U_F$ attacks the particles sent out from Bob and Charlie. In order to induce no error in Step 4, Eve's final probe state should be irrelevant to not only Bob and Charlie's operations but also Alice, Bob and Charlie's measurement results.*

**Proof.** $U_E$ can turn $|0\rangle$ and $|1\rangle$ respectively into

$$U_E(|0\rangle|\chi\rangle_E) = \gamma_{00}|0\rangle|\chi_{00}\rangle + \gamma_{01}|1\rangle|\chi_{01}\rangle, \tag{3}$$

$$U_E(|1\rangle|\chi\rangle_E) = \gamma_{10}|0\rangle|\chi_{10}\rangle + \gamma_{11}|1\rangle|\chi_{11}\rangle, \tag{4}$$

where $|\chi_{00}\rangle, |\chi_{01}\rangle, |\chi_{10}\rangle$ and $|\chi_{11}\rangle$ are the probe states of Eve, $|\gamma_{00}|^2 + |\gamma_{01}|^2 = 1$ and $|\gamma_{10}|^2 + |\gamma_{11}|^2 = 1$.

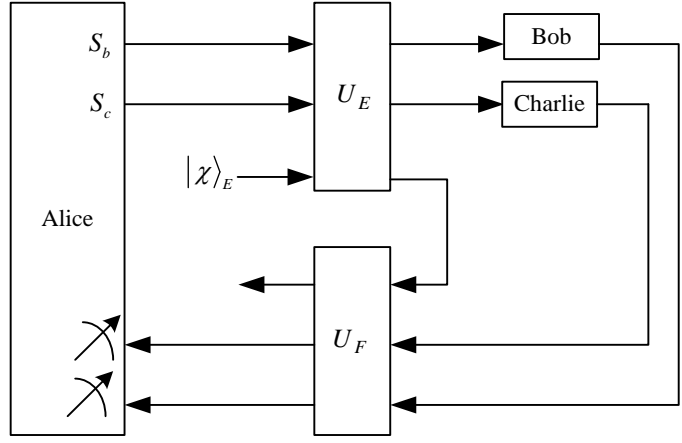

Fig.1 Eve's entangle-measure attack with two unitaries $U_E$ and $U_F$

According to Stinespring dilation theorem, the composite system global state ahead of Bob and Charlie's operation is

$$\begin{aligned}
U_E(|G_1\rangle_{abc}|\chi\rangle_E) &= U_E\left[\frac{1}{2}(|001\rangle+|010\rangle+|100\rangle+|111\rangle)_{abc}|\chi\rangle_E\right] \\
&= \frac{1}{2}\big[|0\rangle_a(\gamma_{00}|0\rangle_b|\chi_{00}\rangle+\gamma_{01}|1\rangle_b|\chi_{01}\rangle)(\gamma_{10}|0\rangle_c|\chi_{10}\rangle+\gamma_{11}|1\rangle_c|\chi_{11}\rangle) \\
&\quad+|0\rangle_a(\gamma_{10}|0\rangle_b|\chi_{10}\rangle+\gamma_{11}|1\rangle_b|\chi_{11}\rangle)(\gamma_{00}|0\rangle_c|\chi_{00}\rangle+\gamma_{01}|1\rangle_c|\chi_{01}\rangle) \\
&\quad+|1\rangle_a(\gamma_{00}|0\rangle_b|\chi_{00}\rangle+\gamma_{01}|1\rangle_b|\chi_{01}\rangle)(\gamma_{00}|0\rangle_c|\chi_{00}\rangle+\gamma_{01}|1\rangle_c|\chi_{01}\rangle) \\
&\quad+|1\rangle_a(\gamma_{10}|0\rangle_b|\chi_{10}\rangle+\gamma_{11}|1\rangle_b|\chi_{11}\rangle)(\gamma_{10}|0\rangle_c|\chi_{10}\rangle+\gamma_{11}|1\rangle_c|\chi_{11}\rangle)\big] \\
&= \frac{1}{2}\big[|0\rangle_a|0\rangle_b|0\rangle_c(\gamma_{00}\gamma_{10}|\chi_{00}\rangle|\chi_{10}\rangle+\gamma_{10}\gamma_{00}|\chi_{10}\rangle|\chi_{00}\rangle) \\
&\quad+|0\rangle_a|0\rangle_b|1\rangle_c(\gamma_{00}\gamma_{11}|\chi_{00}\rangle|\chi_{11}\rangle+\gamma_{10}\gamma_{01}|\chi_{10}\rangle|\chi_{01}\rangle) \\
&\quad+|0\rangle_a|1\rangle_b|0\rangle_c(\gamma_{01}\gamma_{10}|\chi_{01}\rangle|\chi_{10}\rangle+\gamma_{11}\gamma_{00}|\chi_{11}\rangle|\chi_{00}\rangle) \\
&\quad+|0\rangle_a|1\rangle_b|1\rangle_c(\gamma_{01}\gamma_{11}|\chi_{01}\rangle|\chi_{11}\rangle+\gamma_{11}\gamma_{01}|\chi_{11}\rangle|\chi_{01}\rangle) \\
&\quad+|1\rangle_a|0\rangle_b|0\rangle_c(\gamma_{00}^2|\chi_{00}\rangle|\chi_{00}\rangle+\gamma_{10}^2|\chi_{10}\rangle|\chi_{10}\rangle) \\
&\quad+|1\rangle_a|0\rangle_b|1\rangle_c(\gamma_{00}\gamma_{01}|\chi_{00}\rangle|\chi_{01}\rangle+\gamma_{10}\gamma_{11}|\chi_{10}\rangle|\chi_{11}\rangle) \\
&\quad+|1\rangle_a|1\rangle_b|0\rangle_c(\gamma_{01}\gamma_{00}|\chi_{01}\rangle|\chi_{00}\rangle+\gamma_{11}\gamma_{10}|\chi_{11}\rangle|\chi_{10}\rangle) \\
&\quad+|1\rangle_a|1\rangle_b|1\rangle_c(\gamma_{01}^2|\chi_{01}\rangle|\chi_{01}\rangle+\gamma_{11}^2|\chi_{11}\rangle|\chi_{11}\rangle)\big] \\
&= |0\rangle_a|0\rangle_b|0\rangle_c|E_{000}\rangle+|0\rangle_a|0\rangle_b|1\rangle_c|E_{001}\rangle+|0\rangle_a|1\rangle_b|0\rangle_c|E_{010}\rangle+|0\rangle_a|1\rangle_b|1\rangle_c|E_{011}\rangle \\
&\quad+|1\rangle_a|0\rangle_b|0\rangle_c|E_{100}\rangle+|1\rangle_a|0\rangle_b|1\rangle_c|E_{101}\rangle+|1\rangle_a|1\rangle_b|0\rangle_c|E_{110}\rangle+|1\rangle_a|1\rangle_b|1\rangle_c|E_{111}\rangle, \tag{5}
\end{aligned}$$

where $|E_{000}\rangle = \gamma_{00}\gamma_{10}|\chi_{00}\rangle|\chi_{10}\rangle + \gamma_{10}\gamma_{00}|\chi_{10}\rangle|\chi_{00}\rangle$, $|E_{001}\rangle = \gamma_{00}\gamma_{11}|\chi_{00}\rangle|\chi_{11}\rangle + \gamma_{10}\gamma_{01}|\chi_{10}\rangle|\chi_{01}\rangle$, $|E_{010}\rangle = \gamma_{01}\gamma_{10}|\chi_{01}\rangle|\chi_{10}\rangle + \gamma_{11}\gamma_{00}|\chi_{11}\rangle|\chi_{00}\rangle$, $|E_{011}\rangle = \gamma_{01}\gamma_{11}|\chi_{01}\rangle|\chi_{11}\rangle + \gamma_{11}\gamma_{01}|\chi_{11}\rangle|\chi_{01}\rangle$,

$|E_{100}\rangle = \gamma_{00}^2 |\chi_{00}\rangle |\chi_{00}\rangle + \gamma_{10}^2 |\chi_{10}\rangle |\chi_{10}\rangle$ , $|E_{101}\rangle = \gamma_{00}\gamma_{01} |\chi_{00}\rangle |\chi_{01}\rangle + \gamma_{10}\gamma_{11} |\chi_{10}\rangle |\chi_{11}\rangle$ ,

$|E_{110}\rangle = \gamma_{01}\gamma_{00} |\chi_{01}\rangle |\chi_{00}\rangle + \gamma_{11}\gamma_{10} |\chi_{11}\rangle |\chi_{10}\rangle$ and $|E_{111}\rangle = \gamma_{01}^2 |\chi_{01}\rangle |\chi_{01}\rangle + \gamma_{11}^2 |\chi_{11}\rangle |\chi_{11}\rangle$ ].

When Bob receives the particle of $S_b$ sent by Alice, he randomly enters into either the MEASURE-RESEND mode or the REFLECT mode. At the same time, when Charlie receives the particle of $S_c$ sent by Alice, she also randomly enters into either the MEASURE-RESEND mode or the REFLECT mode. Hereafter, Eve performs $U_F$ on the particles sent by Bob and Charlie.

(i) Case ①: both Bob and Charlie enter into the MEASURE-RESEND mode. The composite system global state of Eq.(5) is evolved into $|r\rangle_a |s\rangle_b |t\rangle_c |E_{rst}\rangle$, where $r,s,t \in \{0,1\}$. In order that Eve induces no error in Step 4, for each selected position, Alice's $Z$ basis measurement result on the corresponding particle in her own hand, Bob's $Z$ basis measurement result on the particle of $S_b$ sent by Alice and Charlie's $Z$ basis measurement result on the particle of $S_c$ sent by Alice must be correctly correlated, according to the definition of $|G_1\rangle_{abc}$ described in Eq.(1). Hence, it generates

$$|E_{000}\rangle = |E_{011}\rangle = |E_{101}\rangle = |E_{110}\rangle = \mathbf{0} . \tag{6}$$

Eve performs $U_F$ on the particles sent by Bob and Charlie. In order that Eve induces no error in Step 4, the states of the fresh particles generated by Bob and Charlie mustn't be altered by $U_F$. Hence, it generates

$$U_F \left( |r\rangle_a |s\rangle_b |t\rangle_c |E_{rst}\rangle \right) = |r\rangle_a |s\rangle_b |t\rangle_c |F_{rst}\rangle , \text{ where } r,s,t \in \{0,1\} \text{ and } \overline{r} = s \oplus t . \tag{7}$$

(ii) Case ②: Bob enters into the MEASURE-RESEND mode and Charlie enters into the REFLECT mode. The composite system global state of Eq.(5) is evolved into $|0\rangle_a |0\rangle_b |0\rangle_c |E_{000}\rangle + |0\rangle_a |0\rangle_b |1\rangle_c |E_{001}\rangle + |1\rangle_a |0\rangle_b |0\rangle_c |E_{100}\rangle + |1\rangle_a |0\rangle_b |1\rangle_c |E_{101}\rangle$ when Bob's $Z$ basis measurement result is $|0\rangle_b$ and $|0\rangle_a |1\rangle_b |0\rangle_c |E_{010}\rangle + |0\rangle_a |1\rangle_b |1\rangle_c |E_{011}\rangle + |1\rangle_a |1\rangle_b |0\rangle_c |E_{110}\rangle + |1\rangle_a |1\rangle_b |1\rangle_c |E_{111}\rangle$ when Bob's $Z$ basis measurement result is $|1\rangle_b$.

Firstly, consider that Bob's $Z$ basis measurement result is $|0\rangle_b$. Eve performs $U_F$ on the particles sent by Bob and Charlie. As a result, based on Eq.(6) and Eq.(7), the composite system global state is turned into

$$U_F \left( |0\rangle_a |0\rangle_b |0\rangle_c |E_{000}\rangle + |0\rangle_a |0\rangle_b |1\rangle_c |E_{001}\rangle + |1\rangle_a |0\rangle_b |0\rangle_c |E_{100}\rangle + |1\rangle_a |0\rangle_b |1\rangle_c |E_{101}\rangle \right)$$

$$= |0\rangle_a |0\rangle_b |1\rangle_c |F_{001}\rangle + |1\rangle_a |0\rangle_b |0\rangle_c |F_{100}\rangle$$

$$= \frac{1}{\sqrt{2}} \left( |\psi^+\rangle_{ac} + |\psi^-\rangle_{ac} \right) |0\rangle_b |F_{001}\rangle + \frac{1}{\sqrt{2}} \left( |\psi^+\rangle_{ac} - |\psi^-\rangle_{ac} \right) |0\rangle_b |F_{100}\rangle$$

$$= \frac{1}{\sqrt{2}} |\psi^+\rangle_{ac} |0\rangle_b \left( |F_{001}\rangle + |F_{100}\rangle \right) + \frac{1}{\sqrt{2}} |\psi^-\rangle_{ac} |0\rangle_b \left( |F_{001}\rangle - |F_{100}\rangle \right) . \tag{8}$$

In order that Eve induces no error in Step 4, for each selected position, Alice's $Z$ basis measurement result on the particle from Bob and the Bell basis measurement result on the corresponding particle in her own hand and the particle from Charlie should be correctly related to Bob's $Z$ basis measurement result, according to Eq.(2). Thus, we can derive from Eq.(8) that

$$|F_{001}\rangle = |F_{100}\rangle . \tag{9}$$

Then, consider that Bob's $Z$ basis measurement result is $|1\rangle_b$. Eve performs $U_F$ on the particles sent by Bob and Charlie. As a result, based on Eq.(6) and Eq.(7), the composite system global state is turned into

$$U_F \left( |0\rangle_a |1\rangle_b |0\rangle_c |E_{010}\rangle + |0\rangle_a |1\rangle_b |1\rangle_c |E_{011}\rangle + |1\rangle_a |1\rangle_b |0\rangle_c |E_{110}\rangle + |1\rangle_a |1\rangle_b |1\rangle_c |E_{111}\rangle \right)$$

$$= |0\rangle_a |1\rangle_b |0\rangle_c |F_{010}\rangle + |1\rangle_a |1\rangle_b |1\rangle_c |F_{111}\rangle$$

$$= \frac{1}{\sqrt{2}} \left( |\phi^+\rangle_{ac} + |\phi^-\rangle_{ac} \right) |1\rangle_b |F_{010}\rangle + \frac{1}{\sqrt{2}} \left( |\phi^+\rangle_{ac} - |\phi^-\rangle_{ac} \right) |1\rangle_b |F_{111}\rangle$$

$$= \frac{1}{\sqrt{2}} |\phi^+\rangle_{ac} |1\rangle_b \left( |F_{010}\rangle + |F_{111}\rangle \right) + \frac{1}{\sqrt{2}} |\phi^-\rangle_{ac} |1\rangle_b \left( |F_{010}\rangle - |F_{111}\rangle \right). \tag{10}$$

In order that Eve induces no error in Step 4, for each selected position, Alice's $Z$ basis measurement result on the particle from Bob and the Bell basis measurement result on the corresponding particle in her own hand and the particle from Charlie should be correctly related to Bob's $Z$ basis measurement result, according to Eq.(2). Thus, we can derive from Eq.(10) that

$$|F_{010}\rangle = |F_{111}\rangle. \tag{11}$$

(iii) Case ③: Bob enters into the REFLECT mode and Charlie enters into the MEASURE-RESEND mode. The composite system global state of Eq.(5) is evolved into $|0\rangle_a |0\rangle_b |0\rangle_c |E_{000}\rangle + |0\rangle_a |1\rangle_b |0\rangle_c |E_{010}\rangle + |1\rangle_a |0\rangle_b |0\rangle_c |E_{100}\rangle + |1\rangle_a |1\rangle_b |0\rangle_c |E_{110}\rangle$ when Charlie's $Z$ basis measurement result is $|0\rangle_c$ and $|0\rangle_a |0\rangle_b |1\rangle_c |E_{001}\rangle + |0\rangle_a |1\rangle_b |1\rangle_c |E_{011}\rangle + |1\rangle_a |0\rangle_b |1\rangle_c |E_{101}\rangle + |1\rangle_a |1\rangle_b |1\rangle_c |E_{111}\rangle$ when Charlie's $Z$ basis measurement result is $|1\rangle_c$.

Firstly, consider that Charlie's $Z$ basis measurement result is $|0\rangle_c$. Eve performs $U_F$ on the particles sent by Bob and Charlie. Hence, based on Eq.(6) and Eq.(7), the composite system global state is turned into

$$U_F \left( |0\rangle_a |0\rangle_b |0\rangle_c |E_{000}\rangle + |0\rangle_a |1\rangle_b |0\rangle_c |E_{010}\rangle + |1\rangle_a |0\rangle_b |0\rangle_c |E_{100}\rangle + |1\rangle_a |1\rangle_b |0\rangle_c |E_{110}\rangle \right)$$

$$= |0\rangle_a |1\rangle_b |0\rangle_c |F_{010}\rangle + |1\rangle_a |0\rangle_b |0\rangle_c |F_{100}\rangle$$

$$= \frac{1}{\sqrt{2}} \left( |\psi^+\rangle_{ab} + |\psi^-\rangle_{ab} \right) |0\rangle_c |F_{010}\rangle + \frac{1}{\sqrt{2}} \left( |\psi^+\rangle_{ab} - |\psi^-\rangle_{ab} \right) |0\rangle_c |F_{100}\rangle$$

$$= \frac{1}{\sqrt{2}} |\psi^+\rangle_{ab} |0\rangle_c \left( |F_{010}\rangle + |F_{100}\rangle \right) + \frac{1}{\sqrt{2}} |\psi^-\rangle_{ab} |0\rangle_c \left( |F_{010}\rangle - |F_{100}\rangle \right). \tag{12}$$

In order that Eve induces no error in Step 4, for each selected position, Alice's $Z$ basis measurement result on the particle from Charlie and the Bell basis measurement result on the corresponding particle in her own hand and the particle from Bob should be correctly related to Charlie's $Z$ basis measurement result, according to Eq.(2). Thus, we can derive from Eq.(12) that

$$|F_{010}\rangle = |F_{100}\rangle. \tag{13}$$

Secondly, consider that Charlie's $Z$ basis measurement result is $|1\rangle_c$. Eve performs $U_F$ on the particles sent by Bob and Charlie. Hence, based on Eq.(6) and Eq.(7), the composite system global state is turned into

$$U_F \left( |0\rangle_a |0\rangle_b |1\rangle_c |E_{001}\rangle + |0\rangle_a |1\rangle_b |1\rangle_c |E_{011}\rangle + |1\rangle_a |0\rangle_b |1\rangle_c |E_{101}\rangle + |1\rangle_a |1\rangle_b |1\rangle_c |E_{111}\rangle \right)$$

$$= |0\rangle_a |0\rangle_b |1\rangle_c |F_{001}\rangle + |1\rangle_a |1\rangle_b |1\rangle_c |F_{111}\rangle$$

$$= \frac{1}{\sqrt{2}} \left( |\phi^+\rangle_{ab} + |\phi^-\rangle_{ab} \right) |1\rangle_c |F_{001}\rangle + \frac{1}{\sqrt{2}} \left( |\phi^+\rangle_{ab} - |\phi^-\rangle_{ab} \right) |1\rangle_c |F_{111}\rangle$$

$$= \frac{1}{\sqrt{2}} |\phi^+\rangle_{ab} |1\rangle_c (|F_{001}\rangle + |F_{111}\rangle) + \frac{1}{\sqrt{2}} |\phi^-\rangle_{ab} |1\rangle_c (|F_{001}\rangle - |F_{111}\rangle). \quad (14)$$

In order that Eve induces no error in Step 4, for each selected position, Alice's $Z$ basis measurement result on the particle from Charlie and the Bell basis measurement result on the corresponding particle in her own hand and the particle from Bob should be correctly related to Charlie's $Z$ basis measurement result, according to Eq.(2). Thus, we can derive from Eq.(14) that

$$|F_{001}\rangle = |F_{111}\rangle. \quad (15)$$

So far, it can be derived from Eq.(9), Eq.(11), Eq.(13) and Eq.(15) that

$$|F_{001}\rangle = |F_{010}\rangle = |F_{100}\rangle = |F_{111}\rangle = |F\rangle. \quad (16)$$

(iv) Case ④: both Bob and Charlie enter into the REFLECT mode. The composite system global state of Eq.(5) is kept unchanged before Eve performs $U_F$.

Eve performs $U_F$ on the particles sent by Bob and Charlie. Hence, based on Eq.(5), Eq.(6), Eq.(7) and Eq.(16), the composite system global state is turned into

$$U_F \left[ U_E \left( |G_1\rangle_{abc} |\chi\rangle_E \right) \right]$$

$$= U_F \big( |0\rangle_a |0\rangle_b |0\rangle_c |E_{000}\rangle + |0\rangle_a |0\rangle_b |1\rangle_c |E_{001}\rangle + |0\rangle_a |1\rangle_b |0\rangle_c |E_{010}\rangle + |0\rangle_a |1\rangle_b |1\rangle_c |E_{011}\rangle$$

$$+ |1\rangle_a |0\rangle_b |0\rangle_c |E_{100}\rangle + |1\rangle_a |0\rangle_b |1\rangle_c |E_{101}\rangle + |1\rangle_a |1\rangle_b |0\rangle_c |E_{110}\rangle + |1\rangle_a |1\rangle_b |1\rangle_c |E_{111}\rangle \big)$$

$$= U_F \big( |0\rangle_a |0\rangle_b |1\rangle_c |E_{001}\rangle + |0\rangle_a |1\rangle_b |0\rangle_c |E_{010}\rangle + |1\rangle_a |0\rangle_b |0\rangle_c |E_{100}\rangle + |1\rangle_a |1\rangle_b |1\rangle_c |E_{111}\rangle \big)$$

$$= |0\rangle_a |0\rangle_b |1\rangle_c |F_{001}\rangle + |0\rangle_a |1\rangle_b |0\rangle_c |F_{010}\rangle + |1\rangle_a |0\rangle_b |0\rangle_c |F_{100}\rangle + |1\rangle_a |1\rangle_b |1\rangle_c |F_{111}\rangle$$

$$= \big( |0\rangle_a |0\rangle_b |1\rangle_c + |0\rangle_a |1\rangle_b |0\rangle_c + |1\rangle_a |0\rangle_b |0\rangle_c + |1\rangle_a |1\rangle_b |1\rangle_c \big) |F\rangle$$

$$= |G_1\rangle_{abc} |F\rangle. \quad (17)$$

In order that Eve induces no error in Step 4, for each selected position, Alice's GHZ-like basis measurement result on the corresponding particle in Alice's hand, the particle from Bob and the particle from Charlie should be identical to $|G_1\rangle_{abc}$. This requirement is naturally satisfied, according to Eq.(17).

(v) Inserting Eq.(16) into Eq.(7) generates

$$U_F \left( |r\rangle_a |s\rangle_b |t\rangle_c |E_{rst}\rangle \right) = |r\rangle_a |s\rangle_b |t\rangle_c |F\rangle, \text{ where } r, s, t \in \{0,1\} \text{ and } \bar{r} = s \oplus t. \quad (18)$$

Inserting Eq.(16) into Eq.(8) generates

$$U_F \left( |0\rangle_a |0\rangle_b |0\rangle_c |E_{000}\rangle + |0\rangle_a |0\rangle_b |1\rangle_c |E_{001}\rangle + |1\rangle_a |0\rangle_b |0\rangle_c |E_{100}\rangle + |1\rangle_a |0\rangle_b |1\rangle_c |E_{101}\rangle \right)$$

$$= \sqrt{2} |\psi^+\rangle_{ac} |0\rangle_b |F\rangle. \quad (19)$$

Inserting Eq.(16) into Eq.(10) generates

$$U_F \left( |0\rangle_a |1\rangle_b |0\rangle_c |E_{010}\rangle + |0\rangle_a |1\rangle_b |1\rangle_c |E_{011}\rangle + |1\rangle_a |1\rangle_b |0\rangle_c |E_{110}\rangle + |1\rangle_a |1\rangle_b |1\rangle_c |E_{111}\rangle \right)$$

$$= \sqrt{2} |\phi^+\rangle_{ac} |1\rangle_b |F\rangle. \quad (20)$$

Inserting Eq.(16) into Eq.(12) generates

$$U_F \left( |0\rangle_a |0\rangle_b |0\rangle_c |E_{000}\rangle + |0\rangle_a |1\rangle_b |0\rangle_c |E_{010}\rangle + |1\rangle_a |0\rangle_b |0\rangle_c |E_{100}\rangle + |1\rangle_a |1\rangle_b |0\rangle_c |E_{110}\rangle \right)$$

$$= \sqrt{2} |\psi^+\rangle_{ab} |0\rangle_c |F\rangle. \quad (21)$$

Inserting Eq.(16) into Eq.(14) generates

$$U_F \left( |0\rangle_a |0\rangle_b |1\rangle_c |E_{001}\rangle + |0\rangle_a |1\rangle_b |1\rangle_c |E_{011}\rangle + |1\rangle_a |0\rangle_b |1\rangle_c |E_{101}\rangle + |1\rangle_a |1\rangle_b |1\rangle_c |E_{111}\rangle \right)$$

$$= \sqrt{2}|\phi^+\rangle_{ab}|1\rangle_c|F\rangle. \quad (22)$$

In a word, according to Eqs.(17-22), we can obtain that when Eve imposes her entangle-measure attack with $U_E$ and $U_F$ shown as Fig.1, in order to induce no error in Step 4, Eve's final probe state should be irrelevant to not only Bob and Charlie's operations but also Alice, Bob and Charlie's measurement results.

(3) The double controlled-not (CNOT) attacks

In Step 1, Alice keeps $S_a$ stationary, and sends the particles of $S_b$ to Bob and the particles of $S_c$ to Charlie one by one.

Firstly, we consider single double CNOT attack. Without loss of generality, here take Eve attacking the particle of $S_b$ sent out from Alice for example. Eve imposes the first CNOT attack on the particle of $S_b$ sent out from Alice and her own auxiliary qubit $|0\rangle_E$, where the former and the latter are the control qubit and the target qubit, respectively. The composite system global state after Eve's first CNOT operation can be described as

$$CNOT_{bE}\left(|G_1\rangle_{abc} \otimes |0\rangle_E\right) = \frac{1}{2}\left(|0010\rangle + |0101\rangle + |1000\rangle + |1111\rangle\right)_{abcE}. \quad (23)$$

In Step 2, Bob randomly enters into the REFLECT mode or the MEASURE-RESEND mode for the received particle. At the same time, Charlie randomly enters into the REFLECT mode or the MEASURE-RESEND mode for the received particle. Because Eve doesn't know Bob's operation, in order not to be discovered, Eve has to implement the second CNOT attack, where her auxiliary qubit and the particle from Bob are the target qubit and the control qubit, respectively. In the following, we analyze four different Cases.

Case ①: both Bob and Charlie enter into the MEASURE-RESEND mode. The composite system global state of Eq.(23) is evolved into $|0010\rangle_{abcE}$, $|0101\rangle_{abcE}$, $|1000\rangle_{abcE}$ or $|1111\rangle_{abcE}$. After Eve's second CNOT attack, the composite system global state is evolved into

$$CNOT_{bE}\left(|0010\rangle_{abcE}\right) = |0010\rangle_{abcE}, \quad \text{if Bob and Charlie's measurement results are } |0\rangle_b, |1\rangle_c; \quad (24)$$

$$CNOT_{bE}\left(|0101\rangle_{abcE}\right) = |0100\rangle_{abcE}, \quad \text{if Bob and Charlie's measurement results are } |1\rangle_b, |0\rangle_c; \quad (25)$$

$$CNOT_{bE}\left(|1000\rangle_{abcE}\right) = |1000\rangle_{abcE}, \quad \text{if Bob and Charlie's measurement results are } |0\rangle_b, |0\rangle_c; \quad (26)$$

$$CNOT_{bE}\left(|1111\rangle_{abcE}\right) = |1110\rangle_{abcE}, \quad \text{if Bob and Charlie's measurement results are } |1\rangle_b, |1\rangle_c; \quad (27)$$

Case ②: Bob enters into the MEASURE-RESEND mode and Charlie enters into the REFLECT mode. The composite system global state of Eq.(23) is evolved into $\left(|0010\rangle + |1000\rangle\right)_{abcE}$ or $\left(|0101\rangle + |1111\rangle\right)_{abcE}$. After Eve's second CNOT attack, the composite system global state is evolved into

$$CNOT_{bE}\left(|0010\rangle + |1000\rangle\right)_{abcE} = \sqrt{2}|\psi^+\rangle_{ac}|0\rangle_b|0\rangle_E, \quad \text{if Bob's measurement result is } |0\rangle_b; \quad (28)$$

$$CNOT_{bE}\left(|0101\rangle + |1111\rangle\right)_{abcE} = \sqrt{2}|\phi^+\rangle_{ac}|1\rangle_b|0\rangle_E, \quad \text{if Bob's measurement result is } |1\rangle_b. \quad (29)$$

Case ③: Bob enters into the REFLECT mode and Charlie enters into the MEASURE-RESEND mode. The composite system global state of Eq.(23) is evolved into $\left(|0101\rangle + |1000\rangle\right)_{abcE}$ or $\left(|0010\rangle + |1111\rangle\right)_{abcE}$. After Eve's second CNOT attack, the composite system global state is evolved into

$$CNOT_{bE}\left(|0101\rangle + |1000\rangle\right)_{abcE} = \sqrt{2}|\psi^+\rangle_{ab}|0\rangle_c|0\rangle_E, \quad \text{if Charlie's measurement result is } |0\rangle_c; \quad (30)$$

$$CNOT_{bE}(|0010\rangle+|1111\rangle)_{abcE} = \sqrt{2}|\phi^+\rangle_{ab}|1\rangle_c|0\rangle_E, \quad \text{if Charlie's measurement result is }|1\rangle_c. \quad (31)$$

Case ④: both Bob and Charlie enter into the REFLECT operation. After Eve's second CNOT attack, the composite system global state of Eq.(23) is evolved into

$$CNOT_{bE}^{\otimes 2}(|G_1\rangle_{abc}\otimes|0\rangle_E) = \frac{1}{2}(|0010\rangle+|0100\rangle+|1000\rangle+|1110\rangle)_{abcE} = |G_1\rangle_{abc}|0\rangle_E. \quad (32)$$

It can be derived from Eqs.(24-32) that, although Eve's this kind of double CNOT attacks cannot be detected in Step 4, she has no knowledge about not only Bob and Charlie's operations but also Alice, Bob and Charlie's measurement results, since the final state of her auxiliary particle constantly is $|0\rangle_E$.

Secondly, we consider twice double CNOT attacks. Eve imposes the CNOT attack on the particle of $S_b$ sent out from Alice and $|0\rangle_E$. In the same time, Eve imposes the CNOT attack on the particle of $S_c$ sent out from Alice and her own auxiliary qubit $|0\rangle_F$. Here, the particles of $S_b$ and $S_c$ are the control qubits, while Eve's auxiliary qubits are the corresponding target qubits. The composite system global state after Eve's CNOT operations can be expressed as

$$CNOT_{bEcF}(|G_1\rangle_{abc}\otimes|0\rangle_E\otimes|0\rangle_F) = \frac{1}{2}(|00101\rangle+|01010\rangle+|10000\rangle+|11111\rangle)_{abcEF}. \quad (33)$$

In Step 2, both Bob and Charlie randomly enter into the REFLECT mode or the MEASURE-RESEND mode for the received particles. For not being discovered, Eve has to launch the CNOT attacks again, where the particles from Bob and Charlie are the control qubits, and Eve's auxiliary qubits are the corresponding target qubits. Four different Cases are also discussed in detail.

Case ①: both Bob and Charlie enter into the MEASURE-RESEND mode. The composite system global state of Eq.(33) is turned into $|00101\rangle_{abcEF}$, $|01010\rangle_{abcEF}$, $|10000\rangle_{abcEF}$ or $|11111\rangle_{abcEF}$. The composite system global state after Eve's CNOT attack is turned into

$$CNOT_{bEcF}(|00101\rangle_{abcEF}) = |00100\rangle_{abcEF}, \text{ if Bob and Charlie's measurement results are }|0\rangle_b|1\rangle_c; \quad (34)$$

$$CNOT_{bEcF}(|01010\rangle_{abcEF}) = |01000\rangle_{abcEF}, \text{ if Bob and Charlie's measurement results are }|1\rangle_b|0\rangle_c; \quad (35)$$

$$CNOT_{bEcF}(|10000\rangle_{abcEF}) = |10000\rangle_{abcEF}, \text{ if Bob and Charlie's measurement results are }|0\rangle_b|0\rangle_c; \quad (36)$$

$$CNOT_{bEcF}(|11111\rangle_{abcEF}) = |11100\rangle_{abcEF}, \text{ if Bob and Charlie's measurement results are }|1\rangle_b|1\rangle_c; \quad (37)$$

Case ②: Bob enters into the MEASURE-RESEND mode and Charlie enters into the REFLECT mode. The composite system global state of Eq.(33) is turned into $(|00101\rangle+|10000\rangle)_{abcEF}$ or $(|01010\rangle+|11111\rangle)_{abcEF}$. The composite system global state after Eve's CNOT attack is turned into

$$CNOT_{bEcF}(|00101\rangle+|10000\rangle)_{abcEF} = \sqrt{2}|\psi^+\rangle_{ac}|0\rangle_b|0\rangle_E|0\rangle_F, \text{ if Bob's measurement result is }|0\rangle_b; \quad (38)$$

$$CNOT_{bEcF}(|01010\rangle+|11111\rangle)_{abcEF} = \sqrt{2}|\phi^+\rangle_{ac}|1\rangle_b|0\rangle_E|0\rangle_F, \text{ if Bob's measurement result is }|1\rangle_b. \quad (39)$$

Case ③: Bob enters into the REFLECT mode and Charlie enters into the MEASURE-RESEND mode. The composite system global state of Eq.(33) is turned into $(|01010\rangle+|10000\rangle)_{abcEF}$ or $(|00101\rangle+|11111\rangle)_{abcEF}$. The composite system global state after Eve's CNOT attack is turned into

$$CNOT_{bEcF}(|01010\rangle+|10000\rangle)_{abcEF} = \sqrt{2}|\psi^+\rangle_{ab}|0\rangle_c|0\rangle_E|0\rangle_F, \text{ if Charlie's measurement result is }|0\rangle_c;$$
$$(40)$$

$$CNOT_{bEcF}\left(|00101\rangle+|11111\rangle\right)_{abcEF}=\sqrt{2}|\phi^+\rangle_{ab}|1\rangle_c|0\rangle_E|0\rangle_F, \quad \text{if Charlie's measurement result is }|1\rangle_c.$$
(41)

Case ④: both Bob and Charlie enter into the REFLECT mode. The composite system global state of Eq.(33) after Eve's CNOT attack is turned into

$$CNOT_{bEcF}^{\otimes 2}\left(|G_1\rangle_{abc}\otimes|0\rangle_E\otimes|0\rangle_F\right)=\frac{1}{2}\left(|00100\rangle+|01000\rangle+|10000\rangle+|11100\rangle\right)_{abcEF}=|G_1\rangle_{abc}|0\rangle_E|0\rangle_F. \quad (42)$$

It can be derived from Eqs.(34-42) that, although Eve can hide her trace when launching this kind of double CNOT attacks, she has no way to know not only Bob and Charlie's operations but also Alice, Bob and Charlie's measurement results, since the final states of her auxiliary particles always are $|0\rangle_E|0\rangle_F$.

(4) The measure-resend attack

In Step 1, Alice delivers the particle of $S_b$ to Bob and the particle of $S_c$ to Charlie. Eve uses the $Z$ basis to measure the particle of $S_b$ flying out from the site of Alice and delivers the resulted one to Bob. After Eve's $Z$ basis measurement, the original $|G_1\rangle_{abc}$ is randomly evolved into $|\psi^+\rangle_{ac}|0\rangle_b$ or $|\phi^+\rangle_{ac}|1\rangle_b$. Without losing of generality, assume that the original $|G_1\rangle_{abc}$ is evolved into $|\psi^+\rangle_{ac}|0\rangle_b$. If both Bob and Charlie enter into the MEASURE-RESEND mode, the global state $|\psi^+\rangle_{ac}|0\rangle_b$ will be randomly collapsed into $|001\rangle_{abc}$ or $|100\rangle_{abc}$; and Eve will be perceived with the probability of 0 if this position is selected to perform the security check in Step 4. If Bob enters into the MEASURE-RESEND mode and Charlie enters into the REFLECT mode, the global state $|\psi^+\rangle_{ac}|0\rangle_b$ will kept unchanged; and Eve will also be perceived with the probability of 0 if this position is selected to perform the security check in Step 4. If Bob enters into the REFLECT mode and Charlie enters into the MEASURE-RESEND mode, the global state $|\psi^+\rangle_{ac}|0\rangle_b$ will randomly collapsed into $|001\rangle_{abc}$ or $|100\rangle_{abc}$; and Eve will also be perceived with the probability of $1/2$ if this position is selected to perform the security check in Step 4. If both Bob and Charlie enter into the REFLECT mode, the global state $|\psi^+\rangle_{ac}|0\rangle_b$ will kept unchanged; and Eve will also be perceived with the probability of $1/2$ during the security check in Step 4. In a word, with respect to one particle of $S_b$, Eve can be perceived with the probability of $(1/4)\times(1/2)\times(1/2)+(1/4)\times(1/2)=3/16$. As there are $8(n+\tau)$ particles in $S_b$, Eve's measure-resend attack on these particles can be perceived with the probability of $1-(1-3/16)^{8(n+\tau)}=1-(13/16)^{8(n+\tau)}$, which becomes 1 when is $n$ extremely big.

Similarly, when Eve performs her measure-resend attack on the particles of $S_c$ flying out form the site of Charlie, she is also detected with the probability of $1-(13/16)^{8(n+\tau)}$.

(5) The intercept-resend attack

Eve generates the fake particle within the $Z$ basis, intercepts the particle of $S_b$ flying out from the site of Alice, and delivers the fake one to Bob. Without loss of generality, assume that Eve's fake particle is $|0\rangle_E$. As a result, the global state is evolved into $|G_1\rangle_{abc}|0\rangle_E$. If both Bob and Charlie enter into the MEASURE-RESEND mode, the global state $|G_1\rangle_{abc}|0\rangle_E$ will randomly collapsed into $|\psi^+\rangle_{ab}|00\rangle_{Ec}$ or $|\phi^+\rangle_{ab}|01\rangle_{Ec}$; and Eve will be perceived with the probability of $1/2$ if this position is selected to perform the security check in Step 4. If Bob enters into the MEASURE-RESEND mode and Charlie enters into the REFLECT mode, the global state $|G_1\rangle_{abc}|0\rangle_E$ will be kept unchanged; and Eve will be perceived with the probability of $1/2$ if this position is selected to perform the security check in Step 4. If Bob enters into the REFLECT mode and Charlie enters into the MEASURE-RESEND mode, the global state $|G_1\rangle_{abc}|0\rangle_E$ will randomly collapsed into $(|0010\rangle+|1000\rangle)_{aEbc}$ or $(|0001\rangle+|1011\rangle)_{aEbc}$; and Eve will be perceived with the probability of $3/4$ if this position is selected to perform the security check in Step 4. If both Bob and Charlie enter into the REFLECT mode, the global state $|G_1\rangle_{abc}|0\rangle_E$ will kept unchanged; and Eve will be perceived with the probability of $3/4$ during the security check in Step 4. In a word, with respect to one particle of $S_b$, Eve can be perceived with the probability of $(1/4)\times(1/2)\times(1/2)\times 2+(1/4)\times(1/2)\times(3/4)+(1/4)\times(3/4)=13/32$. As there are $8(n+\tau)$ particles in $S_b$, Eve's measure-resend attack on these particles can be perceived with the probability of $1-(1-13/32)^{8(n+\tau)}=1-(19/32)^{8(n+\tau)}$, which becomes 1 when is $n$ extremely big.

Similarly, when Eve performs her measure-resend attack on the particles of $S_c$ flying out form the site of Charlie, she is also detected with the probability of $1-(1-13/32)^{8(n+\tau)}=1-(19/32)^{8(n+\tau)}$.

## 4 Discussions and conclusions

We use $CE$ of Ref.[4] to evaluate the communication efficiency for the proposed protocol:

$$CE = \frac{LK}{LQ+LC}, \qquad (43)$$

where $LK$, $LQ$ and $LC$ are the length of the private keys established, the number of qubits expended, and the number of classical bits expended, respectively. Note that here ignores the classical bits expended for security checks.

In the proposed protocol, Alice distributes $K_{AB}$ to Bob and $K_{AC}$ to Charlie, and in the meanwhile, Bob and Charlie share $K_A$ together, so $LK=3n$. Alice prepares $8(n+\tau)$ $|G_1\rangle_{abc}$, and sends $S_b$ to Bob and $S_c$ to Charlie; when Bob enters into the MEASURE-RESEND mode for half particles of $S_b$, he is required to prepare $4(n+\tau)$ new particles within the $Z$ basis in the found states; and when Charlie enters into the MEASURE-RESEND mode for half particles of $S_c$, she is also required to prepare $4(n+\tau)$ new particles within the $Z$ basis in the found states; hence, we get $LQ=8(n+\tau)\times 3+4(n+\tau)\times 2=32(n+\tau)$. No classical communication take places, so $LC=0$. In a word, we get $\eta=\frac{3n}{32(n+\tau)}$.

We draw a comparison for the proposed protocol with the present SQKD protocols based on three-particle entangled states in Refs.[21,22,24]. The comparison results are listed in Table 2, where the $X$ basis refers to $\{|+\rangle,|-\rangle\}$. According to Table 2, the proposed protocol takes advantage over the first protocol of Ref.[21] on the aspects of the existence of a third party, the initial quantum resource and the usage of quantum memory for the semiquantum user; the proposed protocol takes advantage over the second protocol of Ref.[21] on the aspects of the existence of a third party, the initial quantum resource, the usage of Hadamard gate and the usage of quantum memory for the semiquantum user; the proposed protocol takes advantage over the second protocol of Ref.[22] on the aspect of the existence of a third party; the proposed protocol takes advantage over the protocol of Ref.[24] on the aspect of the usage of Pauli operations; and the proposed protocol is the only protocol which possesses the functions of both SQKD and SQSS simultaneously.

Table 2 Comparison of the proposed protocol with the present SQKD protocols based on three-particle entangled states

|  | The first protocol of Ref.[21] | The second protocol of Ref.[21] | The first protocol of Ref.[22] | The second protocol of Ref.[22] | The protocol of Ref.[24] | This protocol |
|---|---|---|---|---|---|---|
| Function | Establish a private key between one quantum party and one semiquantum party with a quantum third party | Establish a private key between two semiquantum parties with a quantum third party | Establish a private key between one quantum party and one semiquantum party without a third party | Establish a private key between two semiquantum parties with a quantum third party | Establish a private key between one quantum party and one semiquantum party without a third party | Establish two different private keys between one quantum party and two semiquantum parties respectively, and in the meanwhile, make two semiquantum parties share another private key of the quantum party |
| Whether exists a third party | Yes | Yes | No | Yes | No | No |
| Initial quantum resource | GHZ states and single-qubit states | GHZ states and single-qubit states | GHZ-like states | GHZ-like states | GHZ-like states | GHZ-like states |
| Quantum measurements from the quantum party | The $Z$ basis measurements and the $X$ basis measurements | The $Z$ basis measurements and the $X$ basis measurements | The $Z$ basis measurements and the Bell basis measurements | The $Z$ basis measurements and the GHZ-like basis measurements | The $Z$ basis measurements, the Bell basis measurements and the GHZ-like basis measurements | The $Z$ basis measurements, the Bell basis measurements and the GHZ-like basis measurements |
| Quantum measurements from the classical party | The $Z$ basis measurements | The $Z$ basis measurements | The $Z$ basis measurements | The $Z$ basis measurements | The $Z$ basis measurements | The $Z$ basis measurements |
| Usage of Pauli operations | No | No | No | No | Yes | No |
| Usage of Hadamard gates | No | Yes | No | No | No | No |
| Usage of quantum entanglement swapping | No | No | No | No | No | No |
| Usage of quantum memory for the semiquantum user | Yes | Yes | No | No | No | No |

Moreover, we draw a comparison for the proposed protocol with the present SQSS protocols based on three-particle entangled states in Refs.[27,31,34]. The comparison results are listed in Table 3. According to Table 3, the proposed protocol takes advantage over the first protocol of Ref.[27] on the aspect of the usage of delay lines; and the proposed protocol is the only protocol which possesses the functions of both SQKD and SQSS simultaneously.

In a word, in this paper, in order to accomplish the goal of establishing two different private keys between one quantum party and two semiquantum parties respectively, and making two semiquantum parties share another private key of the quantum party in the meanwhile, we construct a novel hybrid protocol for SQKD and SQSS by using GHZ-like states. This protocol is validated to be immune to Eve's various attacks, including the Trojan horse attacks, the entangle-measure attack, the double CNOT attacks, the measure-resend attack and the intercept-resend attack. To our best knowledge, this protocol is the only protocol possessing the functions of both SQKD and SQSS simultaneously until now.

Table 3 Comparison of the proposed protocol with the present SQSS protocols based on three-particle entangled states

| | The first protocol of Ref.[27] | The second protocol of Ref.[27] | The protocol of Ref.[31] | The protocol of Ref.[34] | This protocol |
|---|---|---|---|---|---|
| Function | Make two semiquantum parties share a private key of the quantum party | Make two semiquantum parties share a private key of the quantum party | Make two semiquantum parties share a private key of the quantum party | Make two semiquantum parties share a private key of the quantum party | Establish two different private keys between one quantum party and two semiquantum parties respectively, and in the meanwhile, make two semiquantum parties share another private key of the quantum party |
| Whether exists a third party | No | No | No | No | No |
| Initial quantum resource | GHZ-like states | GHZ-like states | GHZ-like states | Three-particle W states | GHZ-like states |
| Quantum measurements from the quantum party | The $Z$ basis measurements, the Bell basis measurements and the GHZ-like basis measurements | The $Z$ basis measurements, the Bell basis measurements and the GHZ-like basis measurements | The $Z$ basis measurements, the Bell basis measurements and the GHZ-like basis measurements | The $Z$ basis measurements and the Bell basis measurements | The $Z$ basis measurements, the Bell basis measurements and the GHZ-like basis measurements |
| Quantum measurements from the classical party | The $Z$ basis measurements | The $Z$ basis measurements | The $Z$ basis measurements | The $Z$ basis measurements | The $Z$ basis measurements |
| Usage of Pauli operations | No | No | No | No | No |
| Usage of Hadamard gates | No | No | No | No | No |
| Usage of quantum entanglement swapping | No | No | No | No | No |
| Usage of quantum memory for the semiquantum user | No | No | No | No | No |
| Usage of delay lines | Yes | No | No | No | No |

## Acknowledgments

Funding by the National Natural Science Foundation of China (Grant No.62071430) and the Fundamental Research Funds for the Provincial Universities of Zhejiang (Grant No.JRK21002) is gratefully acknowledged.